\documentclass[onecolumn,amsmath,amssymb,12pt,superscriptaddress,nofootinbib]{revtex4}
\pdfoutput=1

\usepackage[latin1]{inputenc}
\usepackage[english]{babel}
\usepackage{amssymb}
\usepackage{amsmath}
\usepackage{amsthm}
\usepackage[]{graphicx}
\usepackage[]{subfigure}
\usepackage{tensor}
\usepackage{color}
\usepackage{cancel}
\usepackage{setspace}
\usepackage{fancyhdr}
\usepackage{framed}
\usepackage{braket}
\usepackage{float}
\usepackage{tikz}

\newcommand{\be}{\begin{equation}}
\newcommand{\ee}{\end{equation}}
\usepackage[bookmarks,linktocpage, colorlinks=true, plainpages = false, citecolor = blue,  linkcolor=blue, urlcolor = magenta, filecolor = blue]{hyperref} 

\usepackage{mathrsfs}
\usepackage[title]{appendix}
\numberwithin{equation}{section}

\begin{document}

\allowdisplaybreaks

\title{A fresh look at the influence of gravity on the quantum Hall effect}
\vspace{.3in}

\author{F. Hammad}
\email{fhammad@ubishops.ca}
\affiliation{Department of Physics and Astronomy, Bishop's University, 2600 College Street, Sherbrooke, QC, J1M~1Z7
Canada}
\affiliation{Physics Department, Champlain 
College-Lennoxville, 2580 College Street, Sherbrooke,  
QC, J1M~0C8 Canada}
\affiliation{D\'epartement de Physique, Universit\'e de Montr\'eal,\\
2900 Boulevard \'Edouard-Montpetit,
Montr\'eal, QC, H3T 1J4
Canada} 

\author{A. Landry} \email{alexandre.landry.1@umontreal.ca} 
\affiliation{D\'epartement de Physique, Universit\'e de Montr\'eal,\\
2900 Boulevard \'Edouard-Montpetit,
Montr\'eal, QC, H3T 1J4
Canada} 

\author{K. Mathieu} \email{kmathieu17@ubishops.ca}
\affiliation{Department of Physics and Astronomy, Bishop's University, 2600 College Street, Sherbrooke, QC, J1M~1Z7
Canada}

\begin{abstract}
We study the quantum Hall effect inside a gravitational field. First, we review the influence of the gravitational field of the Earth on the quantum Hall effect. Taking the gravitational field of the Earth to be uniform along the vertical direction, we compute the affected quantized Hall resistivity. Then, we investigate how the gravitational field modifies the Landau levels of a particle moving between two massive hemispheres in the presence of a constant and uniform magnetic field perpendicular to the plane of motion. We find that the familiar degeneracy of the Landau levels is removed and the spacing between the latter becomes dependent on the mass density of the hemispheres and on the gravitational constant $G$. We use this result to show that the quantum Hall effect in a thin conductor, sandwiched between two massive hemispheres, should yield a slightly different variation of the Hall resistivity with the applied magnetic field. We then argue that the well-known problem of the gravitationally induced electric field, that might a priori be thought to hinder the effect of gravity, has actually a beneficial role as it amplifies the latter. We finally discuss whether there is a possibility of using the quantum Hall effect to probe the inverse-square law of gravity.
\end{abstract}
\maketitle




\section{Introduction}\label{sec:I}
Two of the praised properties of the quantum Hall effect (QHE) \cite{QHEDiscovery1,QHEDiscovery2} are its independence of the shape of the conductor as well as its insensitivity to small bulk variations of an externally applied voltage and of the small variations due to internal impurities that are within a scale that is below the order of magnitude of the magnetic length \cite{SmoothE,Book1,Oca}. The very manifestation of the effect rests, indeed, on topology and on the presence of disorder in the conductor \cite{Book2,Book1}. All that is required for the effect to arise is a relatively clean conductor, traversed by a direct longitudinal current and submerged --- at low temperatures --- inside a constant and uniform magnetic field that is perpendicular to the flow of the current. That is, the free electrons of the conductor are only subjected to two fields, the constant magnetic and electric fields. The quantized Landau levels, to which the free electrons adhere, together with the presence of a few impurities, are responsible for giving rise to the famous plateaus of the quantized Hall resistivity. In fact, the universality of the effect and the relative uncertainty in the reproducibility of those unique plateaus, up to one part in $10^{10}$ \cite{Metrology2001,Metrology2011}, gave rise immediately after the discovery of the effect to the proposal to use it to measure the fine structure constant \cite{QHEDiscovery1,Discovery}. In view of these remarkable features of the effect and the recent advances in metrology \cite{Metrology2011}, it is tempting to think about other applications of the QHE in fundamental physics, such as to probe the fundamental gravitational interaction.

Recalling that the QHE relies on the presence of a constant electric field in the transverse direction, one cannot help but wonder why the effect would not be influenced by any other constant field acting on the free electrons in the same direction as the transverse electric field. In particular, by submerging the conductor inside a constant and uniform gravitational field, that would be parallel to the electric field, the two fields should simply add up to give a different quantized Hall resistivity than the one expected to appear in the presence of the electric field alone. One can even conceive of a situation in which the transverse electric field is canceled altogether to be replaced by the gravitational field alone. The problem that accompanies such a scenario is, of course, the difficulty of measuring or detecting any variation in the current that would be purely due to the gravitational field alone and using such a current to determine the resistivity as usually done in the QHE. In either scenarios, therefore, the effect would obviously be very small owing to the weakness of the gravitational field. Nevertheless, the mere possibility of the influence of the gravitational field on the QHE makes the latter --- at least in theory --- a real potential tool for probing the gravitational interaction. 

Unfortunately, the influence of the gravitational field on the QHE has attracted very little attention in the literature, with the exception of Ref.\,\cite{Hehl}. In the latter reference, an elegant study of the influence of the gravitational field on the QHE was presented. It was found there that there is no influence of the gravitational field when the latter is perpendicular to the 2-dimensional electron gas (2DEG) inside the conductor, and this up to the order of $\varphi/c^2$, where $\varphi$ is the gravitational potential and $c$ is the speed of light. The influence of a gravitational field that would be parallel to the 2DEG current inside the conductor was found to be of the order of $\varphi/c^2$. The latter specific ratio comes from the relativistic treatment of the interaction of the gravitational field with the electrons \cite{Hehl2,Obukhov}. Our goal in this paper is, first, to review the non-relativistic gravitational influence on the QHE by going back to the case of the uniform gravitational field of the Earth and computing the influence of the latter on the Hall resistivity. Then, we propose a new setup that would allow for a study of the influence of a {\it nonlinear} field of gravity on the electrons' motion in the QHE (see also Ref.\,\cite{Josephson}).

In fact, the uniform gravitational field to use cannot obviously be better than the one provided by the Earth. Indeed, to a very good approximation, the gravitational potential energy of a particle of mass $m$ near the surface of the Earth can be taken to be linear and given by $mgz$ at a given small vertical distance $z$ above a reference point. In particular, it is such a potential that gave rise to the proposal to probe the influence of the gravitational field of the Earth on cold neutrons \cite{UCNinEarth1,UCNinEarth2,UCNinEarth3}. In the usual QHE, one might include inside the Schr\"odinger equation for the free electrons the effect of a constant transverse electric field $E$ in the $z$-direction by adding the potential term $eEz$, where $e$ is the elementary charge. It is then natural to think that the effect of the gravitational field of the Earth could also be taken into account in the QHE by replacing the transverse uniform electric field $E$ by the effective uniform field $E+mg/e$. One is then also naturally tempted to think that this additional contribution could be used to probe the inverse-square law (ISL) via the QHE as the gravitational field of the Earth would be different from $mg$ by correction terms depending on the exact form of the law of gravity. 

This approach, as we shall see, does really introduce an extra correcting term in the Hall resistivity. However, the correction manifest itself as a very small shift in the Hall resistivity. In addition, in order not to overwhelm the gravitational field, as already pointed out in Ref.\,\cite{Hehl}, the magnitude of the electric field involved should not {\it a priori} exceed much $mg/e$, which is of the order of $10^{-10}\,$V/m for electrons and a few orders of magnitude larger for protons and ions. It is then important to seek also a manifestation of the effect of the gravitational field on the QHE at the level of the resistivity plateaus. For that purpose, one needs to gravitationally affect the quantized Landau levels themselves. In fact, it was recently shown that the Landau levels around a massive sphere and around a cylinder are indeed affected by the gravitational field of those massive objects \cite{GravityLandauI,GravityLandauII}. However, the way the levels were found to be modified there was only through the removal of their degeneracy in the orbital quantum number. We propose here a new setup that is capable of affecting gravitationally not only the degeneracy of the Landau levels, but even the spacing between the latter. 

In fact, as was alluded to above, the QHE is insensitive to perturbations in the bulk that are of the order of the magnetic length $\ell_M^2=\hbar/eB$, where $\hbar$ is the reduced Planck constant and $B$ is the strength of the magnetic field. In addition, the very manifestation of the QHE relies heavily on both the spacing between the Landau levels and the degeneracy of each one of them. In Refs.\,\cite{GravityLandauI,GravityLandauII}, it was found that the gravitational field due to a (static and/or rotating) sphere as well as that due to a long cylinder both remove the degeneracy of the Landau levels by splitting the orbital energies of a charged particle going around such massive objects. As we shall see, using, instead, two massive hemispheres separated by a small distance gives rise in the motion of the free 2DEG inside a conductor sandwiched between the two hemispheres to a gravitationally induced harmonic oscillator. Consequently, a setup based on two weakly separated hemispheres not only removes the degeneracy of the Landau levels, but modifies also the spacing between the latter. Furthermore, we show that for a very small separation between the two hemispheres, the deviation of the gravitational potential from that of a pure harmonic oscillator varies, in fact, over distance scales much larger than the magnetic length $\ell_M$. This implies that the QHE should be affected while remaining immune to such deviations. 

Finally, it is well known that when a conductor is immersed inside a gravitational field, the latter is itself always accompanied by an opposite induced electric field acting on the free electrons of the conductor in the same direction as does the gravitational field \cite{GravityInducedE1966,GravityInducedE1967,GravityInducedEThroughIons}. The magnitude of the induced electric field is of the order $\sim0.1\,Mg/e$, where $M$ is the mass of the ions in the conductor's lattice \cite{GravityInducedE1966,GravityInducedE1967,GravityInducedEThroughIons,Hehl}. Unlike in the experiments \cite{FreeFall1,FreeFall2} testing the equivalence principle through the free fall of charged particles, however, we show that the gravitationally induced electric field does not hinder the effect of gravity on the QHE. On the contrary, we show that such an induced electric field actually amplifies the gravitational influence on the QHE. 

The rest of this paper is organized as follows. In Sec.\,\ref{sec:II}, we examine the effect on the QHE of the linear gravitational field provided by the Earth and compute the explicit expression of the modified Hall resistivity. In Sec.\,\ref{sec:III}, we examine the fate of Landau levels of particles moving in a horizontal plane between two close massive hemispheres in the presence of a uniform and constant magnetic field that is normal to the equatorial plane of the hemispheres. We then study how, in turn, this affects the QHE that would arise in a circular conductor --- a Corbino-like disk --- put in between the two hemispheres with the center of the latter coinciding with the center of the conductor. In Sec.\,\ref{sec:IV}, we discuss the effect of the gravitationally induced electric field in both cases, inside the gravitational field of the Earth and then in between the two hemispheres. For both cases we provide real and detailed quantitative predictions with a discussion on the technological limitations (centered around the cryogenics) that presently prohibit the manifestation of some those predictions. In Sec.\,\ref{sec:V}, we discuss whether there is a possibility of using the QHE for testing the ISL. We conclude this paper with a short summary section.

\section{The QHE inside Earth's gravitational field}\label{sec:II}
\begin{figure}[H]
    \centering
    \includegraphics[scale=0.7]{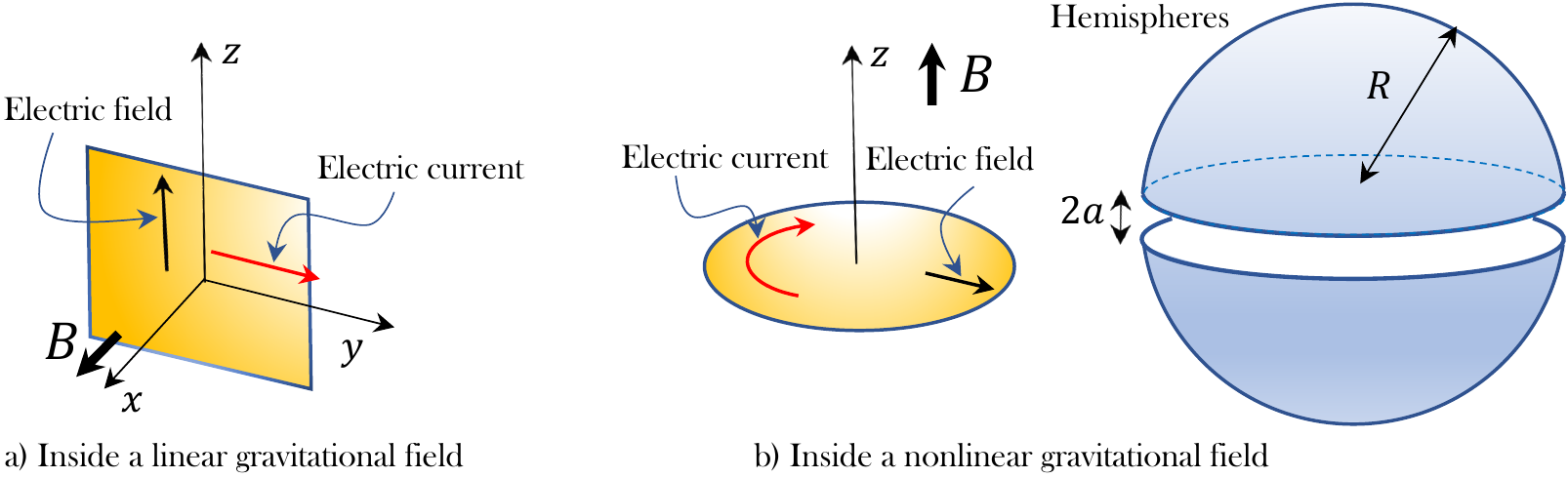}
    \caption{a) A rectangular conductor inside the vertical gravitational field of the Earth. b) A circular conductor, on the left, to be inserted between the two massive hemispheres on the right.}
    \label{fig:Hemispheres}
\end{figure}

The observed resistivity plateaus in the QHE are due to the combination of the quantized Landau levels of the free electrons of the conductor and the few defects of the conductor around which the electrons move. Suppose the conductor containing the 2DEG lies in the vertical $yz$-plane as shown in Figure\,\ref{fig:Hemispheres}-a).

{}
The free electrons of mass $m$ and at distance $z$ from the bottom of the conductor acquire then the gravitational potential energy $mgz$. Suppose also that a constant transverse electric field $E$, acting between the two edges of the conductor, is chosen to be parallel to the $z$-direction and pointing upwards. We assume the constant and uniform magnetic field $B$ is perpendicular to the conductor and parallel to the $x$-direction as in Figure\,\ref{fig:Hemispheres}-a). We adopt the Landau gauge in which the vector potential reads ${\bf A}=(0,-Bz,0)$. As the 2DEG is freely moving parallel to the $y$-direction with momentum $\hbar k$, the wavefunctions will take the simple form $e^{iky}\psi_n(z-z_0)$. These are indeed the eigenfunctions of the electron Hamiltonian $H(z)$, which takes the form,
\begin{align}\label{EarthHamiltonian}
H(z)&=\frac{1}{2m}\left[(p_y-eBz)^2+p_z^2\right]+(eE+mg)z\nonumber\\
&=\frac{p_z^2}{2m}+\frac{1}{2}m\omega_c^2(z-z_0)^2+(eE+mg)z_0+\frac{\left(eE+mg\right)^2}{2m\omega_c^2}.
\end{align}
We have introduced in the second line the cyclotron frequency, defined by $\omega_c=eB/m$, as well as the shifted centers $z_0=k\ell_M^2-m(eE+mg)\ell_M^4/\hbar^2$ along the $z$-direction for each delocalized wavefunction of momentum $\hbar k$. In fact, we recognize in the second line of expression (\ref{EarthHamiltonian}) the Hamiltonian of a simple harmonic oscillator displaced from the origin by the amount $z_0$. The quantized energy eigenvalues of such a Hamiltonian are then given by,
\begin{equation}\label{Eignvalues}
\mathcal{E}_n=\hbar\omega_c\left(n+\frac{1}{2}\right)+(eE+mg)z_0+\frac{\left(eE+mg\right)^2}{2m\omega_c^2}.
\end{equation}
We thus see that the spacing $\hbar\omega_c$ between two consecutive Landau levels is not affected by the gravitational field. However, the shifted centers $z_0$, conditioned by the gravitational field, do influence the current corresponding to a delocalized state $e^{iky}\psi_n(z-z_0)$ in the $y$-direction. In fact, the current $I_y$ in the $y$-direction, given by $-e\braket{v_y}=e\braket{\hbar k-eBz}/m$ for a single moving electron, does acquire a contribution from the gravitational term $mg$ because of the fact the the plane waves are centered at $z_0$; that is, $\braket{z}=z_0$. More explicitly, the current density due to all the free electrons of the sample of surface area $S$ evaluates to,
\begin{equation}\label{EarthCurrent}
J_y=\frac{e}{mS}\sum_{n,k}\braket{\psi_n|\hbar k-eBz|\psi_n}=\frac{\nu N_0e}{S}\left(\frac{eE+mg}{eB}\right)=\left(1+\frac{mg}{eE}\right)\frac{\nu e^2}{h}E.
\end{equation}
Here, $\nu$ is the number of filled Landau levels and $N_0$ is the number of states in the conductor sharing the same Landau level. For a conducting sample of surface area $S$, this degeneracy is given in terms of the quantum of flux $\Phi_0=h/e$ by the ratio $N_0=SB/\Phi_0$ \cite{Book2}.

According to Eq.\,(\ref{EarthCurrent}), then, the Hall resistivity, given by $\rho_{yz}=E/J_y$, exhibits the correction factor $(1+mg/eE)^{-1}$ in front of the familiar quantized expression $h/\nu e^2$. This correction factor depends on the ratio $mg/eE$. As discussed in the Introduction, this ratio is indeed very small. The correction factor brought to the quantized Hall resistivity can thus be taken, to a good approximation, to be $(1-mg/eE)$. For an electric field of the order of $1\,$V/m, we find by plugging the value of the mass $m$ of the electron and using the gravitational acceleration at the surface of the Earth, $g=9.8\,$m/s$^2$, that the ratio $mg/eE$ becomes of the order of $10^{-10}$. This estimate is based on the assumption that the sensitivity to the Hall voltage allowed by the equipment used in the setup is of the order of a few millivolts for a sample size of the order of a millimeter. Such voltage sensitivity is presently easily achievable. Moreover, we see from Eq.\,({\ref{EarthCurrent}}) that any higher sensitivity in measuring the Hall voltage would greatly increase the sensitivity to the gravitational effect on the Hall resistivity as well. Indeed, with a sensitivity on the Hall voltage down to the microvolt within very low temperatures, the transverse electric field $E$ could be as low as  $10^{-3}$V/m. With such a low electric field, the correcting factor $mg/eE$ accompanying the Hall resistivity would be increased to attain the order of $10^{-7}$. Obviously, if it were not for technological challenges, one can always create strong artificial gravitational fields, thanks to the weak equivalence principle, by putting the setup inside a centrifugal machine\footnote{Many thanks to Valerio Faraoni for suggesting to use our local centrifuge.}. Gravitational accelerations as high as $10^5g$ might then easily be achieved for which the correcting factor $mg/eE$ in front of the quantum Hall resistivity would reach the order of $10^{-2}$. 

As we shall see in detail in Sec.\,\ref{sec:IV}, however, by taking into account the gravitationally induced electric field in the conductor, the small correction term due to Earth's gravity, obtained here by ignoring such an induced electric field, becomes actually much larger. Before we do that, however, we are first going to examine in the next section a new configuration of the gravitational field obtained with two very close massive hemispheres. Such a setup gives rise to a nonlinear gravitational field which allows the spacing between the Landau levels themselves, as well as their degeneracy, to be gravitationally affected.

\section{The QHE between two massive hemispheres}\label{sec:III}
The key feature in using two hemispheres is the appearance of a gravitationally induced simple harmonic oscillator that combines with the magnetically induced one to give rise to gravitationally modified Landau levels. The latter would, in turn, modify the shape of the plateaus of the quantized Hall resistivity. For this purpose, we first need to find the gravitational potential at any point in between the two hemispheres at a distance $r$ away from the center of mass of the two hemispheres. The detailed calculations of the gravitational potential based on (i) the purely Newtonian potential, (ii) on the Newtonian potential plus a Yukawa-like correction and (iii) on a Newtonian potential plus a power-law correction are all given in Appendix \ref{A}. Before we use those various expressions of the potential, however, we shall first briefly describe here the physical setup and then study the Schr\"odinger equation that governs the motion of a free electron inside the conductor (or any charged particle for that matter) when such a gravitational interaction is taken into account.
\subsection{Landau levels between two massive hemispheres}\label{LandauHemisphere}
The setup simply consists of a circular conductor (on the left in Figure \ref{fig:Hemispheres}-b)) to be sandwiched between the two massive grounded hemispheres (on the right in Figure \ref{fig:Hemispheres}-b)). The disk, along which the current flows, is subjected to a transverse radial constant electric field $E$. The magnetic field, being perpendicular to the plane of the conductor, its corresponding vector potential $\bf A$ can be described in the cylindrical coordinates $(r,\phi,z)$ by adopting the symmetric gauge in which it reads, ${\bf A}=(0,\frac{1}{2}Br,0)$. 

Let $V_g(r)$ denote the gravitational potential felt by an electron of mass $m$ moving in the equatorial plane sandwiched between the two hemispheres. The motion of the electron being planar around the center, we choose the ansatz $\psi(r,\phi)=e^{i\ell\phi}R(r)$ for the wavefunction of the particle; $\ell$ being a positive integer. The Hamiltonian in the cylindrical coordinates reads $H(r,\phi)=({\bf p}+e{\bf A})^2/2m+mV_g(r)$. The Schr\"odinger equation for an electron of energy $\mathcal{E}$ then takes the form,
\begin{equation}\label{Schrodinger}
\frac{{\rm d}^2R}{{\rm d}r^2}
+\frac{1}{r}\frac{{\rm d}R}{{\rm d}r}+\Bigg(\frac{2m\mathcal{E}}{\hbar^2}+\frac{eB\ell}{\hbar}-\frac{e^2B^2r^2}{4\hbar^2}-\frac{\ell^2}{r^2}-\frac{2m^2V_g(r)}{\hbar^2}\Bigg)R=0.
\end{equation}
The solution to this equation without the gravitational potential $V_g(r)$ has been given in Refs.\,\cite{GravityLandauI,GravityLandauII} with all the steps outlined in great detail. The result is a function $R(r)\sim r^{\ell} e^{-\frac{eB}{4\hbar}r^2}$, with the proportionality factor being a hypergeometric function. More important, however, is the fact that the energy eigenvalues are those of a simple harmonic oscillator $\mathcal{E}_n=\hbar\omega_c(n+\frac{1}{2})$, with the cyclotron frequency $\omega_c$ given again by $\omega_c=eB/m$ \cite{LandauLifshitz}. 

Now, all the gravitational potentials found in Appendix \ref{A} have the mathematical form $\frac{1}{2}Kr^2$ (with $K$ a constant with the dimensions of a squared angular frequency) which is that of the potential of a simple harmonic oscillator, to which adds weaker deviation terms, which we shall denote here by $C_g(r)$. This perturbing term $C_g(r)$ exhibits various higher-than-three powers of the ratios $r/R$ and $a/R$ between, respectively, the distance $r$ of the electrons from the center and the radius $R$ of the hemispheres and the separation distance $2a$ and $R$. The constant $K$ depends on the gravitational constant $G$ and on the mass density of the hemispheres, but its specific form varies depending on the law one adopts for the gravitational interaction. For an ISL, for a Yukawa-like deviation from the ISL and for a power-law deviation from the ISL, the constant $K$ is given, respectively, by Eqs.\,(\ref{KNew}), (\ref{KNew+Yuk}) and (\ref{KPower}) below. As a consequence, Eq.\,(\ref{Schrodinger}) can then be written in the following more useful form:
\begin{equation}\label{SchroSuggestive}
\frac{{\rm d}^2R}{{\rm d}r^2}
+\frac{1}{r}\frac{{\rm d}R}{{\rm d}r}+\Bigg[\frac{2m\mathcal{E}}{\hbar^2}+\frac{eB\ell}{\hbar}-\frac{(e^2B^2+4m^2K)r^2}{4\hbar^2}-\frac{\ell^2}{r^2}-\frac{2m^2C_g(r)}{\hbar^2}\Bigg]R=0.
\end{equation}
In order to solve this equation and extract the quantization condition on the energy $\mathcal{E}$, we proceed along the same steps used in Ref.\,\cite{GravityLandauI} to solve the simpler equation (\ref{Schrodinger}) without the last term. Here, we are going to solve Eq.\,(\ref{SchroSuggestive}) without the last term which just brings in a very weak perturbation. 

Let us chose the ansatz $R(r)=e^{-\frac{\beta}{4}r^2}r^{\ell}v(r)$, where $v(r)$ is a function of $r$ only and $\beta=\sqrt{e^2B^2+4m^2K}/\hbar$. Inserting this expression of $R(r)$ into Eq.\,(\ref{SchroSuggestive}), and denoting by a prime a derivative with respect to the variable $r$, the equation becomes,
\begin{align}\label{vEquation}
&r v''(r)+\left(2\ell+1-\beta r^2\right)v'(r)+(\alpha-\beta)r\, v(r)=0.\nonumber\\
&\alpha=\frac{2m\mathcal{E}}{\hbar^2}+\left(\frac{eB}{\hbar}-\beta\right)\ell.
\end{align}
By performing the change of variable $z=\tfrac{1}{2}\beta r^2$, the differential equation in $v(r)$ takes the following canonical form \cite{BookKummer},
\begin{equation}\label{Canonical}
zv''(z)+\left(\ell+1-z\right)v'(z)-\left(\frac{1}{2}-\frac{\alpha}{2\beta}\right)v(z)=0.
\end{equation}
This equation is a confluent hypergeometric differential equation \cite{BookConfluent}. Its general solution is a linear combination of two confluent hypergeometric functions $_1F_1(a;b;z)$, also known as Kummer's functions \cite{BookKummer}. Keeping only from the linear combination the term that converges at the origin $r=0$, we get the following expression for the radial function $R(r)$ \cite{GravityLandauI}:
\begin{equation}\label{Solution}
R(r)=A\,r^{\ell} e^{-\frac{\beta}{4}r^2}\,_1F_1\left(\frac{1}{2}-\frac{\alpha}{2\beta};\ell+1;\frac{\beta}{2}r^2\right).
\end{equation}
We have introduced here the constant of integration $A$ that also plays the role of a normalization constant \cite{GravityLandauI}. This expression, in turn, is diverging exponentially for large $r$ because of the Kummer function. For this reason, one should impose the following condition on the first argument of the latter \cite{BookKummer,BookConfluent}:
\begin{equation}\label{Quantization}
\frac{1}{2}-\frac{\alpha}{2\beta}=-n.    
\end{equation}
Here, $n$ is a non-negative integer for which the confluent hypergeometric function becomes then indeed a finite-degree polynomial in $r^2$. By substituting into this condition the values of $\alpha$ and $\beta$ we defined above, we arrive at the following quantization condition for the energy $\mathcal{E}$ of the electron inside the conductor:
\begin{equation}\label{LandauGravity}
\mathcal{E}_{n\ell}=\hbar\varpi_c\left(n+\frac{1}{2}\right)+\frac{\hbar\ell}{2}\left(\varpi_c-\omega_c\right). 
\end{equation}
Here, we introduced the modified cyclotron frequency,
\begin{equation}\label{Varomega}
\varpi_c=\sqrt{\frac{e^2B^2+4m^2K}{m^2}}=\omega_c\sqrt{1+\frac{4m^2K}{e^2B^2}},    
\end{equation}as opposed to the original cyclotron frequency $\omega_c=eB/m$. The first term in the result (\ref{LandauGravity}) represents the usual form of the quantized Landau energy levels. The second term, however, depends on the orbital quantum number $\ell$. The consequence of having such an extra term is the removal of the usual familiar infinite degeneracy of Landau levels as the energy of the latter acquires a different value for each different orbital $\ell$. For a small gravitational term $4m^2K$ compared to the magnetic term $e^2B^2$, the following first-order approximation of Eq.\,(\ref{LandauGravity}) is valid,
\begin{equation}\label{LandauGravityFirstOrder}
\mathcal{E}_{n\ell}\approx\hbar \omega_c\left(1+\frac{2K}{\omega_c^2}\right)\left(n+\frac{1}{2}\right)+\frac{\hbar K\ell}{\omega_c}. 
\end{equation}

Now, this result is valid not only for an electron inside a conductor sandwiched between the two hemispheres, but also for {\it any} charged particle that would happen to be moving between the two hemispheres under the influence of the magnetic and gravitational fields. The effect of the correction term to the Landau levels and the effect of the last term that splits each level in Eq.\,(\ref{LandauGravityFirstOrder}) would therefore both become accentuated for more massive charged particles. In fact, heavy ions and molecules would multiply the correction term to the Landau levels by a factor of at least of the order of $10^{8}$ for an atom of atomic number $6$, like carbon, given that the proton mass is about $1836$ times larger than the electron mass. Similarly, the Landau levels-splitting term is then multiplied by a factor of at least of the order of $10^4$. In fact, given that the cyclotron frequency of the carbon ion is $\omega_c\approx8\times10^{6}\,$rad/s, the unsplit lowest Landau level is at $2\times10^{-7}\,$eV under a magnetic field strength of $1\,$T. The sub-level $\ell\sim10^{19}$ is then found at $\,10^{-7}$eV because, as we shall see later on, the constant $K$ is of the order of $6\times10^{-6}\,$rad$^2$/s$^2$ for a massive sphere of platinum. Thus, by amplifying the splitting of the Landau levels by using molecules which are much heavier, one might be able to resolve between the different possible forms the parameter $K$ can take for the different possible corrections to the Newtonian ISL for gravity one considers.  
Furthermore, in obtaining the result (\ref{LandauGravityFirstOrder}), we have not taken into account the contribution of the weaker perturbing terms gathered inside the term $C_g(r)$ in Eq.\,(\ref{SchroSuggestive}). Such terms would introduce additional, albeit tiny, splittings of the Landau levels. 

With this general result at hand, we can now apply it to the electrons in a conductor to examine its consequences on the QHE.
\subsection{Consequences on the QHE}
As we saw below Eq.\,(\ref{EarthCurrent}), a longitudinal current in the conductor is proportional to the available number $N_0$ of degenerate states at each Landau level as well as to the number $\nu$ of those levels that are filled by the conducting electrons. As for the degeneracy, we just saw that it is removed by the gravitational field between the two hemispheres. As a consequence, the electrons contributing to the total conductance will be spread, according to their energies, over all the sub-levels available at each allowed Landau level. It is actually easy to estimate the new number of the sub-levels $N$ corresponding to each principle Landau level $n$ when $\ell$ is very large. In fact, by setting $R(r)=r^{-1/2}\chi(r)$, Eq.\,(\ref{SchroSuggestive}) takes the following form:
\begin{equation}\label{SchrodLike}
\frac{-\hbar^2}{2m}\chi''+\Bigg[\left(\frac{e^2B^2+4m^2K}{8m}\right)r^2+\frac{\hbar^2(\ell^2-\tfrac{1}{4})}{2mr^2}-\frac{\hbar eB\ell}{2m}+mC_g(r)\Bigg]\chi=\mathcal{E}\chi.
\end{equation}
Apart from the perturbing term $mC_g(r)$, the unperturbed potential $V(r)$ inside the square brackets may be expanded in powers of $r$ around the equilibrium position $r_0$. The latter is found by solving ${\rm d}V/{\rm d}r=0$. For very large orbitals, $\ell\gg1$, we find,
\begin{equation}\label{r_0}
r_0^2\approx\frac{2\hbar\ell}{\sqrt{e^2B^2+4m^2K}}=\frac{2\hbar\ell}{m\varpi_c}.
\end{equation}
Using this expression, the effective potential in Eq.\,(\ref{SchrodLike}) can be approximated by that of a perturbed simple harmonic oscillator, $V_{\rm eff}(r)=\frac{1}{2}m\varpi_c^2(r-r_0)^2+\frac{1}{2}\hbar\ell\left(\varpi_c-\omega_c\right)+mC_g(r)$. The wavefunctions become then concentrated around the approximate radial distance $r_0$. For a circular conductor of surface area $S$, the number $N$ of states comprised inside the radius $r_0$ can then be estimated to be,
\begin{equation}\label{Degeneracy}
    N\approx\frac{S\sqrt{e^2B^2+4m^2K}}{h}\approx N_0\left(1+\frac{2K}{\omega_c^2}\right),
\end{equation}
where, $N_0=SB/\Phi_0$ is the number of states sharing the same Landau principle level $n$ in the absence of the gravitational field.

We thus see that the number of sub-levels available around a given Landau level $n$ is the same as the number of degenerate states emerging in the absence of the gravitational field. This number is in fact only corrected by the ratio $2K/\omega_c^2$. More important, however, is that this number of states is huge. Therefore, given that, in addition, each orbital $\ell$ provides a slightly different energy from another adjacent orbital $\ell+1$, two adjacent principle Landau levels $n$ and $n+1$ do not remain sharply distinct as they are in the case when the gravitational field is absent. In fact, in the latter case, the separation between adjacent Landau levels is simply $\hbar \omega_c$. With the removal of the degeneracy by the gravitational field, however, the sharp Landau levels broaden as the available energies that arise right above a given Landau level exhibit a nearly continuous spectrum as can be seen from Eq.\,(\ref{LandauGravityFirstOrder}). However, due to the finite surface of the conductor, the number of available sub-levels between two consecutive Landau levels is not infinite, but given instead by Eq.\,(\ref{Degeneracy}). As a result, according to Eq.\,(\ref{LandauGravityFirstOrder}) we can estimate the maximum energy gap between two adjacent broadened Landau levels $\mathcal{E}_{n+1}$ and $\mathcal{E}_{n}$ (which is the difference between the lowest sub-level of $\mathcal{E}_{n+1}$ and the highest sub-level of $\mathcal{E}_n$) to be,
\begin{equation}\label{Distance}
\Delta\mathcal{E}\approx\hbar\omega_c\left(1+\frac{2K}{\omega^2_c}\right)-\frac{\hbar KN}{\omega_c}\approx\hbar\omega_c\left(1-\frac{m^2KS}{h eB}\right).
\end{equation}

The consequence of this shrinking of the distance between the broadened Landau levels could only show up in the length of the plateaus of the quantum Hall resistivity without affecting the vertical distance between the plateaus. Indeed, in analogy to what we did for the gravitational field of the Earth in Sec.\,\ref{sec:II}, we can also derive an approximate expression for the azimuthal current density $J_C$ in the circular conductor. Since equation (\ref{SchrodLike}) can be approximated by that of a perturbed simple harmonic oscillator in the radial direction for large $\ell$, we have the following approximation for the Hamiltonian in the presence of a constant radial electric field $E$:
\begin{equation}\label{CorbinoSHO}
H(r)=\frac{p_r^2}{2m}+\frac{1}{2}m\varpi_c^2(r-r_*)^2+\frac{\hbar\ell}{2}\left(\varpi_c-\omega_c\right)\\+eEr_0-\frac{e^2E^2}{2m\varpi_c^2}+mC_g(r).
\end{equation}
Here, $p_r$ stands for $-i\hbar\partial_r$. Also, we have introduced here the shifted center $r_*=r_0-eE/m\varpi_c^2$ along the radial direction for each delocalized wavefunction of linear momentum $\hbar k$ and angular momentum $\hbar\ell=\hbar kr$. Thus, we find, up to the first order in $K/\omega_c^2$, the following approximate estimate for the current density along the circular conductor,
\begin{align}\label{HemisphereCurrent}
J_C&=\frac{e}{mS}\sum_{n,k}\braket{\psi_{n}|\hbar k-\tfrac{1}{2}eBr|\psi_{n}}\approx\frac{\nu Ne^2E}{mS}\frac{\omega_c}{\varpi_c^2}\approx\left(1-\frac{2K}{\omega_c^2}\right)\frac{\nu e^2}{h}E.
\end{align}
In the third step, we have used the expression (\ref{Degeneracy}) of the estimated number of sub-levels $N$ and discarded second- and higher-orders in $K/\omega_c^2$ and $eE/m\omega_c^2$. The integer $\nu$ is the number of filled Landau levels. Thus, the variation of the quantized Hall resistivity with the strength of the magnetic field should display again the characteristic jumps marked by an integer $\nu$ as observed in the absence of the gravitational field. In addition, the magnitude of each jump is again given by the familiar constant $h/e^2$, as in the absence of the gravitational field, up to the negligible correction $2K/\omega_c^2$. This negligible correction to the von Klitzing constant as compared to the correction brought by the gravitational field of the Earth can be understood by the fact that the transverse gravitational field created by the hemispheres is at least $20$ orders of magnitude smaller than that of the Earth. 

The remarkable difference as compared to the effect of the Earth, however, is that each sharp and distinct degenerate Landau level is now split into sub-levels, and is thus broadened according to Eq.\,(\ref{LandauGravityFirstOrder}), and --- to a lesser extent as we saw --- the distance between adjacent groups of split Landau levels shrinks according to Eq.\,(\ref{Distance}). Consequently, as the magnetic field is gradually increased, the 2DEG is provided with a huge number of sub-levels $N_0\sim10^{12}$ per unit area of the sample that can be occupied before the group of such sub-levels becomes inaccessible again and lies above the Fermi level. It is as if the role used to be played by the impurities inside the conductor, which are responsible for creating the mobility gap and giving rise to the horizontal plateaus in the quantum Hall effect, is now also played by the Landau orbitals that provide sub-levels for the 2DEG to occupy. In contrast to the mobility gap created by the localized states made available by the impurities, however, the sub-levels made available by splitting the Landau levels provide extended (delocalized) states for the 2DEG. As such, the filling of these sub-levels is accomplished by the electrons participating in the conduction. As a result, if we increase the electron density in the system or reduce the magnetic field, so that the Fermi level lies within the region of the extended states, we do gain current-carrying states. Therefore, the conductivity should increase at that point. This then should show up in the deformation of the transition from one plateau to another resulting in a more gentle jump than the one observed in the absence of the gravitational field (see Fig. \ref{Fig:Plateaus} b)). In fact, as in the absence of the gravitational field, the localized states provided by the impurities will still make the split Landau levels broader, resulting in the usual observed plateaus. We shall analyse this point in more detail below.

Note also that in our present discussion, the perturbing terms gathered inside the term $mC_g(r)$ in Eq.\,(\ref{CorbinoSHO}) are merely treated as bulk variations of the effective potential felt by the 2DEG. That perturbing potential, as we shall see below, brings, in fact, deviations from the pure simple harmonic oscillator that are of the order of $r^4/R^2$ and higher, where $r$ is the distance of the electron from the center and $R$ is the radius of the two hemispheres. Therefore, keeping in mind that the QHE effect is insensitive to the bulk perturbations that vary over distances much larger than the magnetic length $\ell_M$ \cite{SmoothE,Book1}, we conclude that we may safely take these perturbing terms to be a mere bulk perturbation.

Now, formula (\ref{Distance}) predicts only a shrinking of the distance between adjacent groups of split Landau levels that is of the order of $10^{-14}$ for a one-Tesla magnetic field per unit area of the sample. However, as we shall see in the next section, this correction becomes amplified by a factor of nearly five orders of magnitude. In addition, we shall see that the splitting of the Landau levels will become much more accentuated due to this amplification that the resistivity plateaus will become much more affected than what one gets without taking into account such an induced electric field.

\section{The effect of the induced electric field}\label{sec:IV}
As mentioned in the Introduction, the gravitational field induces an electric field in the conductor which, in turn, affects the free electrons of the latter. A detailed study of such an effect is conducted in Refs.\,\cite{GravityInducedE1966,GravityInducedE1967,GravityInducedEThroughIons}. The simplest way to understand the origin of such an induced electric field is to notice that the atoms making the lattice of the conductor get compressed by the gravitational field due to their own weight. In order to preserve charge neutrality, an electric field in the opposite direction to the gravitational field is then induced by the shifted ions, attracting thus the electrons. In this section, we are first going to review the simple model that allows one to find the induced electric field in the presence of the gravitational field of the Earth. We then examine the effect of this electric field on the quantum Hall resistivity. Next, we adapt the same model to find the induced electric field inside the conductor caused by the nonlinear gravitational field we found between the two hemispheres. We then take into account such an induced field to study the behaviour of the new quantum Hall resistivity.

A simple model to arrive at the induced electric field $E^I$ for the case of the constant gravitational field of the Earth in the $z$-direction is given in Ref.\,\cite{GravityInducedEThroughIons}. The model is based on the balance between, on the one hand, the gravitational and electric forces, $n_0(eE^I+mg)$, acting on the electron gas of equilibrium density $n_0$ and, on the other hand, the gradient of the pressure of the electron gas, $\partial_z p_e$. Then, for an electron gas of density $n$ the average energy is $\varepsilon\propto n^{2/3}$ and, assuming that the electrons' pressure is $p_e=\frac{2}{3}n\varepsilon\propto\frac{2}{3}n^{5/3}$, the balance equation reads, $\frac{10}{9}\varepsilon\,\partial_zn=-n_0(eE^I+mg)$ \cite{GravityInducedEThroughIons}. Next, the atoms of the lattice, of mass $M$, display an atomic density that obeys the following equilibrium equation under the influence of the gravitational field: $C\partial_zn/n_0=-n_0Mg$, where the constant $C$ depends on the elastic properties of the conductor \cite{GravityInducedEThroughIons}. Finally, as the electrons density is conditioned by the atomic density (to preserve charge neutrality), we deduce, by comparing the two previous balance equations, that
$E^I=g(\gamma M-m)/e$, where the constant $\gamma$ is given by $\frac{10}{9}\varepsilon n_0/C$ \cite{GravityInducedEThroughIons}. It is found that for the case of copper the constant $\gamma$ is of the order of $1/7$, so that the contribution of the electron mass $m$ to the induced electric field is negligible. Therefore, the induced electric field reduces to $E^I\approx \frac{1}{7}Mg/e$ \cite{GravityInducedEThroughIons}. 

\subsection{Earth's effect} Going back now to the result (\ref{EarthCurrent}), and adding the contribution $eE^I$ of this induced electric field to the 2DEG, side by side with the effect of the gravitational field, we arrive at the following modification in the quantum Hall resistivity:
\begin{equation}\label{EarthWithInducedE}
J_y=\left(1+\frac{Mg}{7eE}+\frac{mg}{eE}\right)\frac{\nu e^2}{h}E\approx\left(1+\frac{Mg}{7eE}\right)\frac{\nu e^2}{h}E.
\end{equation}
In the second step, we have kept the leading term proportional to the atomic mass of the lattice. Thus, we see that the effect of the gravitational field has been amplified thanks to the compression of the ions of the lattice. As the mass of the ions of the conductor (here, copper) is nearly five orders of magnitude larger than that of the electrons, we have a huge amplification factor indeed. This allows one to go from a gravitational correction of the order of $10^{-10}$ coming from the ratio $mg/eE$, when one does not take into account the induced electric field, to a correction that is four orders of magnitude higher when including the induced field.

In order to better appreciate this effect, let us compute the precise correction the gravitational field of the Earth brings to the quantum Hall resistivity by injecting actual values into formula (\ref{EarthWithInducedE}). First, using the charge of the electron $e$ and Planck's constant $h$, we easily compute the ratio $e^2/h$ to be $3.874\,045\,865\,5\times10^{-5}\,$S, the inverse of which is the famous von Klitzing constant, a measurable resistance quantum of $R_K\equiv25\,812.807\,45...\,\rm \Omega$ \cite{CODATA}. Each of the plateaus in the plot showing the variation of resistivity in the QHE as a function of the magnetic field in the absence of gravity (see the plot on the left in Fig.\,\ref{Fig:Plateaus} below) represents thus an integer multiple of the ratio $h/e^2$. Taking into account the effect of Earth's gravitational field, formula (\ref{EarthWithInducedE}) predicts that each of these plateaus should be shifted down by a constant factor. For a conducting sample made of copper we have the atomic mass $M=1.0552061\times10^{-25}\,$kg. Then, taking the gravitational acceleration at the surface of the Earth to be $g=9.81\,$m/s$^2$ and having a transverse electric field $E=1\,$V/m, the multiplicative factor in Eq.\,(\ref{EarthWithInducedE}) evaluates to $1+9.23\times10^{-7}$. Thereby, the plot for the quantum Hall resistivity should display transverse conductivity plateaus separated by the distance $3.874\,049\,441\,2\times10^{-5}\,$S. This, in turn would translate into an apparent von Klitzing ``constant" $\tilde{R}_K=25\,812.783\,63...\,\rm \Omega$. 

Now, one might argue that since this effect is just an overall upward shift of the Hall conductivity (or, equivalently, a downward shift of the von Klitzing constant), no such difference could ever be detected. However, it is very important to keep in mind that while the correcting factor displayed by formula (\ref{EarthWithInducedE}) is explicitly independent of the magnetic field $B$, it does actually depend on the transverse electric field $E$ as it is inversely proportional to the latter. It is for this reason that we put the word ``constant" between quotation marks when referring to the apparent value $\tilde{R}_K$. In other words, the correction to the quantum Hall resistivity becomes larger for smaller transverse electric fields $E$ and smaller for larger fields. This very interesting and remarkable fact is simply absent in the absence of a transverse gravitational field. By carefully measuring the Hall voltage to a high accuracy, one is then certainly able to detect such a dependence. Thereby, by managing to lower the Hall voltage to the order of a microvolt in a millimeter-size sample, the transverse electric field could decrease to $10^{-3}\,$V/m (and to $10^{-4}\,$V/m in a centimeter-size sample). With such a small value of the transverse electric field, the correcting factor in formula (\ref{EarthWithInducedE}) increases to $1+9.23\times10^{-4}$. The quantized QHE conductivity then increases to $3.877\,621\,609\,8\times10^{-5}\,$S. This yields a relatively dramatic decrease in the apparent von Klitzing constant to $\tilde{R}_K=25\,789.004\,20...\,\rm\Omega$.

It is worth emphasizing here the importance of having the electric field act transversely on the used sample in the QHE experiment. In fact, the remarkable quantitative effect we arrived at above using formula (\ref{EarthWithInducedE}) was possible only because the gravitational field acts parallel to the transverse electric field, and hence the effects of the former add up to the effects of the latter. This key point is related to one of the two intuitively unexpected outcomes of the effect of gravity on the QHE. The first unexpected outcome is related to the fact that the QHE is naturally insensitive to the latitude at which the experiment is conducted on Earth. In this case, while the gravitational potential does indeed vary with the height, it nevertheless remains uniform in the transverse direction. In contrast, in our case the sample is set vertically so that the gravitational potential due to Earth varies linearly in the transverse direction exactly as does the Hall voltage. The second unexpected outcome, to which we turn to in the next subsection, is also related to this point in that it still involves a transverse gravitational field. The only difference, is that the latter is, in addition, nonlinear.

\subsection{The Hemispheres' effect}
Let us now adapt the same model for the case of the nonlinear gravitational field between the two massive hemispheres. Again, the compression of the atoms of the conductor by the gravitational field will be responsible for inducing an electric field in the opposite direction. As the gravitational potential $V_g(r)$ now varies with the radial distance $r$ from the center of mass of the two hemispheres, the induced electric field $E^I$ should be radial as well and should also vary accordingly with the distance $r$. Therefore, the balance equation between the gravitational and electric forces acting on the electron gas and the gradient of the pressure of the latter now reads in the cylindrical coordinates, $\frac{10}{9}\varepsilon\,\partial_rn=-n_0\left[eE^I(r)+m\partial_r V_g(r)\right]$. Similarly, under the influence of the gravitational field the atomic density obeys the equilibrium equation, $C\partial_rn/n_0=-n_0M\partial_rV_g(r)$. As a consequence, by following the same steps as we did for the Earth's gravitational field, the radially induced electric field $E^I(r)$ should be given by
$E^I(r)\approx \frac{1}{7}M\partial_rV_g(r)/e$. As we saw above, the gravitational potential between the two hemispheres has the form $\frac{1}{2}Kr^2+C_g(r)$. This means that the induced electric potential $V_e^I(r)$ inside the conductor is given by $V_e^I(r)\approx \frac{1}{7}M\left[\frac{1}{2}Kr^2+C_g(r)\right]/e$. As such, the cyclotron frequency $\varpi_c$ obtained in the presence of gravity for the modified Landau levels (\ref{LandauHemisphere}) becomes modified into the following induced cyclotron frequency:
\begin{equation}\label{InducedCyclotron}
\varpi_c^I=\sqrt{\frac{e^2B^2}{m^2}+\frac{4M}{7m}K+4K}\approx\omega_c\left(1+\frac{2MK}{7m\omega_c^2}\right).
\end{equation}
In the second step, we have again taken into account the fact that $M/m\gg1$ and discarded the last term inside the square root. The effect of gravity on the cyclotron frequency becomes thus much amplified compared to what we find when not taking into account the induced electric field. 

The Newtonian gravitational potential between the hemispheres, found in Eq.\,(\ref{HemisphereNew}) of Appendix \ref{A}, displays a constant $K$ given by,
\begin{equation}\label{KNew}
K=\frac{4\pi G\rho}{3}\left(1-\frac{3 a}{2R}\right).
\end{equation}
For two one meter-radius hemispheres made of platinum, this constant $K$ is of the order of $10^{-6}\,$rad$^2$/s$^2$. The correction to the cyclotron frequency $\omega_c\sim10^{11}\,{\rm rad/s}$, obtained under a constant magnetic field $B=1$\,T, will therefore be of the order of $10^{-24}$. This correction is obviously negligible. However, what is important, as we saw in Sec.\,\ref{sec:III} and as we shall see shortly, is not the correction to the cyclotron frequency as given by Eq.\,(\ref{InducedCyclotron}) but rather the modification brought to the degeneracy of the Landau levels. Before, we examine the fate of such a degeneracy, let us first look at the fate of the distance between the split Landau levels. 

We saw in Eq.\,(\ref{Distance}) that the distance between two adjacent Landau levels is affected by the gravitational field between the two hemispheres. Taking into account the induced electric potential $V_e^I(r)$, the distance $\Delta\mathcal{E}$ we found in Eq.\,(\ref{Distance}) becomes,
\begin{equation}\label{DistanceInduced}
\Delta\mathcal{E}\approx\hbar\omega_c\left(1-\frac{mMKS}{7h eB}\right).
\end{equation}
For the same size platinum hemispheres used above, and for the same magnitude of $1\,$T for the magnetic field $B$, the order of magnitude of the correcting factor inside the parentheses is $\sim10^{-9}$ per unit area of the sample. This is a larger correction compared to what we get when not taking into account the induced electric field, but it is still way smaller than the correction we found for the current density in the Earth's gravitational field. Furthermore, such a shrinking in the distance between the split Landau levels decreases for larger strengths of the magnetic field. Yet, this shrinking should certainly affect the length of the horizontal plateaus of the quantum Hall resistivity. We shall examine this point in more detail using actual values shortly below.

Going back now to expression (\ref{LandauGravityFirstOrder}), and inserting the induced electric field, we find the following modified formula for the splitting of the energy levels, 
\begin{equation}\label{LandauGravityFirstOrderInduced}
\mathcal{E}_{n\ell}\approx\hbar \omega_c\left(1+\frac{2mMK}{7e^2B^2}\right)\left(n+\frac{1}{2}\right)+\frac{\hbar MK\ell}{7eB}. 
\end{equation}
In order to be able to detect any eventual influence of the magnetic field on the shape of the Hall resistivity plateaus, we have expressed in this formula the correcting term as well as the levels-splitting term as functions of the magnetic field $B$ instead of the cyclotron frequency $\omega_c$ as done in Eq.\,(\ref{LandauGravityFirstOrder}). First, expression (\ref{LandauGravityFirstOrderInduced}) clearly implies that the initially broadened Landau levels, due to the sub-states made available by the defects of the sample, become even more broadened due to the amplification of the levels-splitting by the induced electric field --- compare the distribution of the density of states shown in parts a) and b) of Figure\,\ref{Fig:Plateaus} below. Second, as the ratio $\frac{\hbar MK\ell}{7eB}$ is inversely proportional to the strength of the magnetic field $B$, Eq.\,(\ref{LandauGravityFirstOrderInduced}) entails that the spectrum of sub-states made of individual orbitals $\ell$ becomes closer to a continuous spectrum for stronger magnetic fields. As such, we expect that the effect of gravity on the horizontal plateaus and the steps would be much cleaner for weaker magnetic fields. See Figure \ref{Fig:Plateaus}.
\begin{figure}[H]
    \centering
    \includegraphics[scale=0.7]{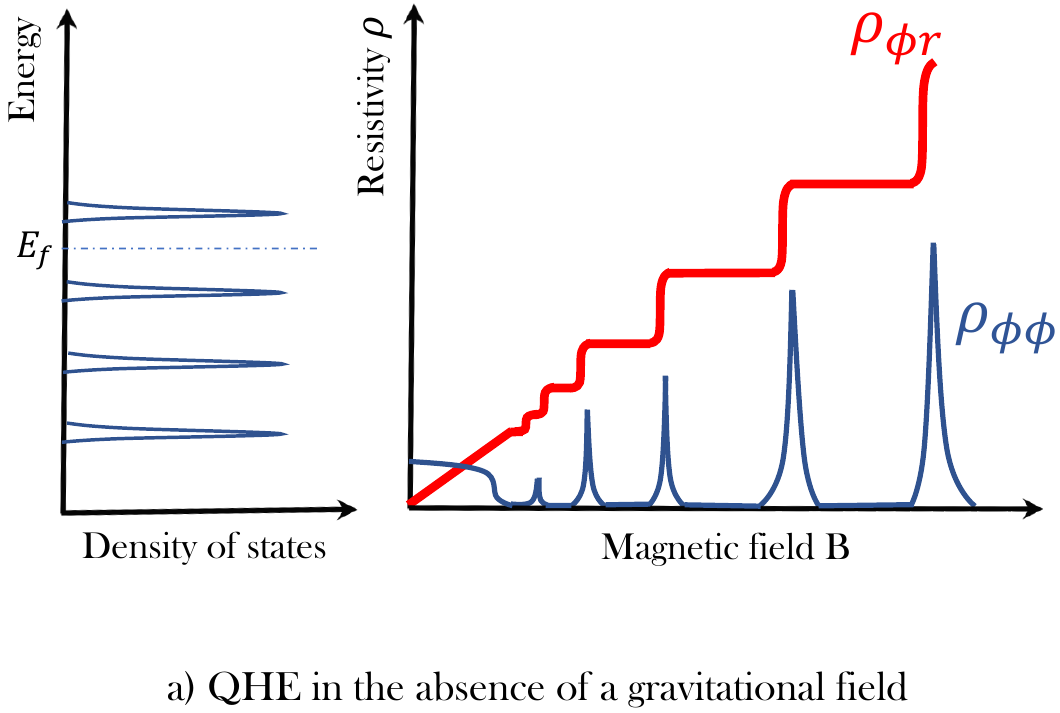}\qquad\quad\includegraphics[scale=0.7]{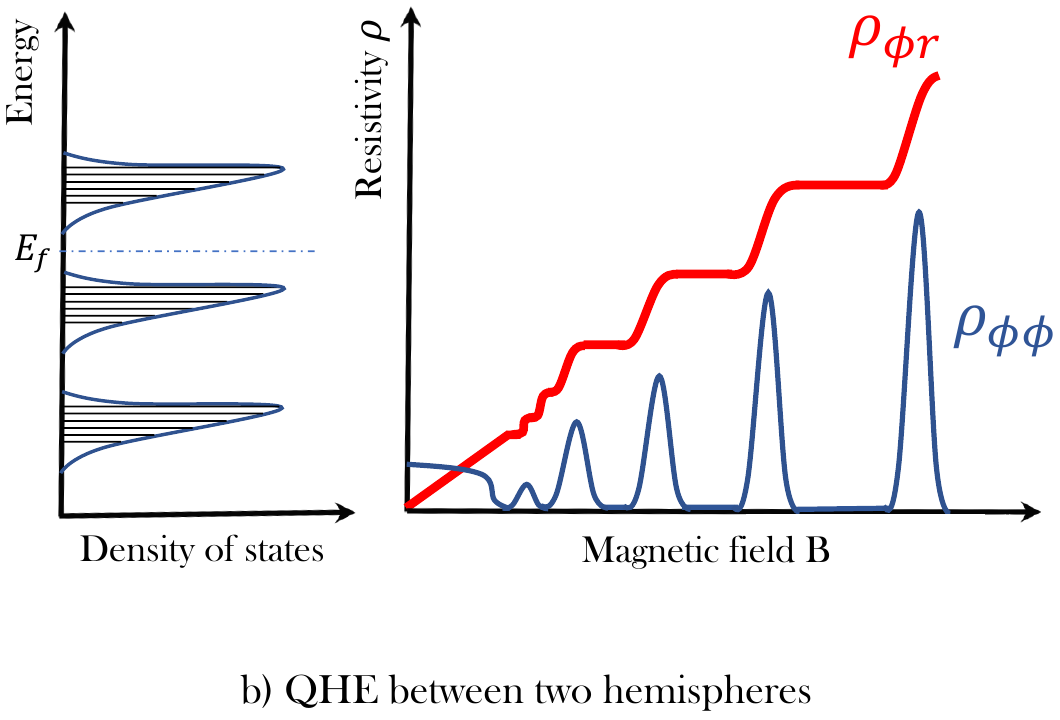}
    \caption{a) The electron's density of state around the Fermi energy $E_f$ and the Hall resistivity $\rho_{\phi r}$ and the azimuthal resistivity $\rho_{\phi\phi}$ inside a circular conductor in the absence of a gravitational field. b) The electron's density of states around the Fermi energy $E_f$ (the horizontal lines represent the split Landau levels) and the Hall resistivity $\rho_{\phi r}$ and the azimuthal resistivity $\rho_{\phi\phi}$ in a circular conductor between two massive hemispheres. The features of both cases are exaggerated in order to see the qualitative difference between the two.}
    \label{Fig:Plateaus}
\end{figure}
Let us now perform a more precise quantitative evaluation of the effect of the massive hemispheres on the QHE by injecting actual values in our formulas (\ref{DistanceInduced}) and (\ref{LandauGravityFirstOrderInduced}). First, let us assume that the massive hemispheres have each a radius $R=1\,$m and that both are made of pure platinum of mass density $\rho=21447\,$kg/m$^3$, separated by a distance $2a=2\times10^{-3}\,$m. Then, the squared angular frequency of the gravitationally induced harmonic oscillator as given by Eq.\,(\ref{KNew}) evaluates to $K\approx6\times10^{-6}\,$rad$^2$/s$^2$. Next, for a conducting sample made of copper we have the atomic mass $M\approx1\times10^{-25}\,$kg. This then yields the value $8\times10^{-14}$ for the correction factor inside the parentheses in Eq.\,(\ref{DistanceInduced}) under a magnetic field strength of $1\,$T and for a sample of surface area $S=1\,$cm$^2$. Recalling now that $\omega_c=eB/m$, formula (\ref{DistanceInduced}) entails that the length of the Hall plateau parallel to the $B$-axis of the graph should shrink at the magnetic field strength of $1\,$T by an amount of the order $\delta B\sim10^{-13}\,$T. Although this is a very small value, it is well-known that the use of a superconductor quantum interference device (SQUID) at very low temperatures does presently allow to reach very high precision measurements of the magnetic field strength variation of the order of $10^{-15}\,$T (see e.g. Ref.\,\cite{SQUID}).

This is not the end of the story, however. In fact, it is already clear from formula (\ref{LandauGravityFirstOrderInduced}) that both the location of the Landau levels and the degeneracy of the latter depend on the applied magnetic field. The first parentheses of the first term in Eq.\,(\ref{LandauGravityFirstOrderInduced}) represent the correction to each Landau level. It is clear, though, that such a correction is too small to make any noticeable difference in our present discussion. A rough estimate shows indeed that the correcting term inside the parentheses is of the order of $10^{-24}$. With the second term, however, the phenomenology becomes more interesting. That term represents the splitting of each Landau level. It is the term responsible for the gravitational broadening of the Landau levels. The width of such a gravitationally induced broadening is found by identifying the largest orbital number $\ell$ with the number $N\approx N_0=SB/\Phi_0$. We then find the following width,
\begin{equation}\label{Width}
    \Delta_g\approx\frac{MKS}{14\pi}.
\end{equation}
With the above values of $M$ and $K$, we find a gravitationally-induced broadening of the Landau levels of the order of $\Delta_g\sim10^{-17}\,$eV for a sample size of $1\,$cm$^2$. Now, while this is still too small to make any difference compared to the thermal broadening $\Delta_{\rm th}\sim k_BT\approx10^{-5}\,$eV under a temperature of $1\,$K, it does make a difference in the quantum Hall plateau-plateau transitions. To see this, we need a brief digression to recall the well-known and actively investigated scaling behaviour of the quantum Hall plateau-plateau transitions.

\begin{figure}[H]
\centering
\includegraphics[scale=0.35]{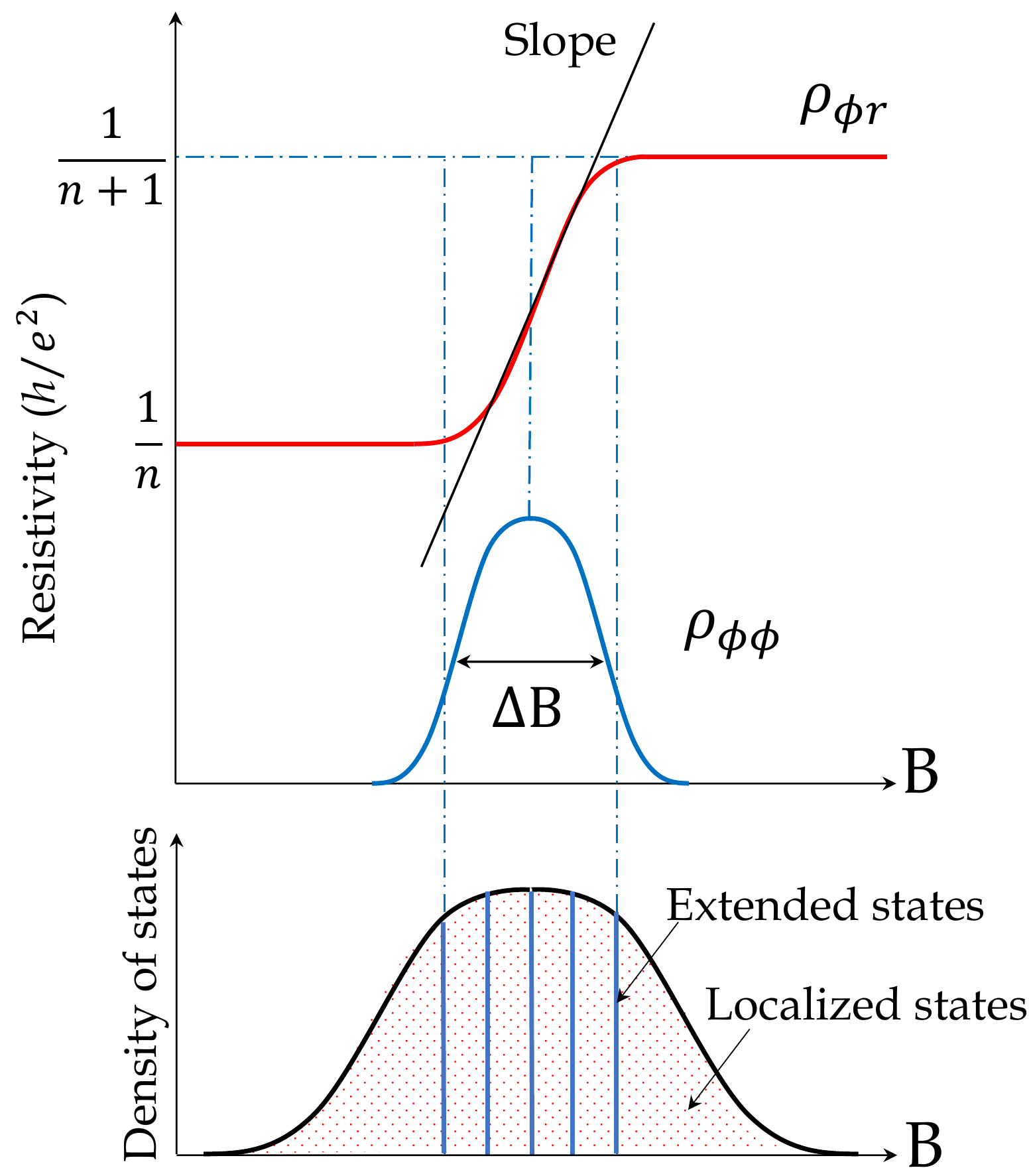}
\caption{a) The top graph shows the variation of the transverse resistivity $\rho_{\phi r}$ and the longitudinal resistivity $\rho_{\phi\phi}$ as functions of the applied magnetic field $B$. The slope of the transition between plateaus scales with temperature like $T^{-\mu}$, with constant $\mu$, for $T\rightarrow0$. In the presence of the nonlinear gravitational field between the massive hemispheres, the slope becomes inversely proportional to $B^2$. Similarly, the width $\Delta B$ of the $\rho_{\phi\phi}$-plot that scales like $T^{\mu}$ for $T\rightarrow0$ becomes inversely proportional to $B$ inside the massive hemispheres. The broadening of the Landau levels, shown by the bottom graph, is due to (i) the gravitationally-induced splitting (vertical lines housing the extended states) and (ii) local defects housing the localized states.}
\label{Fig:Slope}
\end{figure}

In a defect-free sample at a temperature $T=0$, each Landau level gives rise to extremely sharp density of states filled at the Fermi energy by extended electronic states whose wavefunction is delocalized throughout the sample. The presence of defects and finite temperatures changes this picture in the following way. First, the presence of defects in the sample dramatically broadens such a density of states around the Landau levels. This impurity-broadening effect is carried out by the localized electronic states. As such, the longitudinal resistivity remains unchanged as the extended states are unaffected, whence the large horizontal plateaus. Therefore, if it were only for this impurity-induced broadening of the density of states, a sharp discontinuity from one plateau to another would be witnessed. Instead, what is experimentally observed is a smooth transition. Such a smoothness is due to the thermal broadening of the density of states. In fact, as thermal fluctuations around each sharp Landau level affect the extended electronic states near the Fermi energy, the transverse and longitudinal resistivities are accordingly affected as a dissipation (or a metallic regime, i.e. a phase transition) then emerges \cite{PhaseTransition1}. It was found experimentally that as $T\rightarrow0$, the transition phase in InGaAs-InP is characterized by the scaling behaviour $\partial\rho_{xy}/\partial B\propto T^{-\mu}$. The exponent $\mu$ is found to be $\mu\sim0.4$ \cite{PhaseTransition2} (see also Refs.\,\cite{Scaling1991,Scaling2009,Scaling2019} and the review paper \cite{ScalingReview}). The divergence in this power-law scaling at low temperatures translates into a plateau-plateau transition characterized by a slope (see Fig.\,\ref{Fig:Slope}) that is becoming steeper with decreasing temperatures, whence the sharpness of the transitions at very low temperatures. Similarly, it is found that the magnetic width $\Delta B$ of the $\rho_{xx}$-peaks is exponentially deceasing at very low temperatures according to the scaling law $\Delta B\propto T^{\mu}$ \cite{PhaseTransition2}. 

In the presence of the non-linear gravitational field between the hemispheres, however, the relative gravitationally induced broadening of a Landau level is given by $\Delta_g/\hbar\omega_c=m\Delta_g/\hbar eB$. This ratio dictates how much the resistivity departs from a sudden jump. Therefore, using Eq.\,(\ref{Width}) we deduce that by taking into account the gravitationally induced plateau-plateau transitions the slope of the QHE graph in those regions should not display any singular behavior at very low temperatures $T\rightarrow0$ because gravity then takes over and the slope adopts the following form,
\begin{equation}\label{Slope}
\frac{\partial\rho_{xy}}{\partial B}\propto-\frac{mMKS}{7heB^2}.
\end{equation}
Similarly, according to Eq.\,(\ref{LandauGravityFirstOrderInduced}) the magnetic field width $\Delta B$ of the longitudinal resistivity peaks will not decay exponentially anymore as $T\rightarrow0$ since it then takes, independently of the magnetic field, the following universal form,
\begin{equation}\label{RhoPeaks}
\Delta B\approx\frac{mMKS}{7he}.
\end{equation}
With a magnetic field strength of $1\,$T and the values we adopted above for $M$, $K$ and $S$, the slope is shifted from the vertical direction by the amount $\sim10^{-13}\times \tilde{R}_K/\,{\rm T}\sim2.6\times10^{-9}\,{\rm \Omega}/$T and the magnetic width of the $\rho_{\phi\phi}$-peak is of the order $\Delta B\sim10^{-13}\,$T; the same order of magnitude by which the quantum Hall plateau shrinks as we saw above. It is clear, of course, that these minute corrections become meaningful only at very low temperatures that are presently far from being accessible. In fact, the gravitational broadening $\Delta_g$ becomes of the same order as the thermal broadening $\Delta_{\rm th}$ for temperatures of the order of a pico-kelvin. It is worth noting, however, that as temperature is decreased the transport time $\tau$ of the electrons increases and the condition $\omega_c\tau>1$ becomes satisfied for lower magnetic fields \cite{LowFields}, and hence a larger effect on the slope results according to formula (\ref{Slope}) even before reaching such low temperatures.

Finally, we would like to comment now on the various approximations we made in deriving our results. Given that all our approximations have been made based on the fact that the term $mC_g(r)$ in Eqs.\,(\ref{SchrodLike}) and (\ref{CorbinoSHO}) --- which consists of terms of order $\mathcal{O}(r^4/R^4,r^2a^2/R^4)$ and higher (see Appendix \ref{A}) --- has been treated as a perturbation, one might wonder whether adding the amplification term $\frac{1}{7}M/m$ to the latter would prevent it from being a mere perturbation. The answer depends on the ratios $r/R$ and $a/R$ between, respectively, the radial distance $r$ of the 2DEG from the center and the radius $R$ of the two hemispheres, and the separation distance $2a$ and $R$. As the ratio $M/m$ is of the order of $10^6$ for copper, we need the ratios $r/R$ and $a/R$ to be at most of the order of $10^{-4}$. That is, for two one-meter radius hemispheres, the radius $r$ and the separation distance $2a$ between the two hemispheres should not be larger than a few fractions of a millimeter. In addition, for such scales the variation of the perturbation $mC_g(r)$ is still guaranteed to be larger than the magnetic length $\ell_M$. In fact, in this case, we also have $C_g^{-1}(r)\partial_rC_g(r)$ of the order of a fraction of a millimeter which is much larger than the magnetic length $\ell_M\sim250\,$\r{A} under a one-Tesla magnetic field. Moreover, we have $mC_g(r)\ll \hbar\omega_c$, as well as $m\partial^2_rC_g(r)\ll m\omega_c^2$ which testifies about the very slow variation of the bulk perturbation.

Whenever the radius $r$ from the center exceeds a few fractions of a millimeter, the term $mC_g(r)$ cannot be considered as a mere perturbation anymore. Then, the ``protective'' feature that the QHE is not affected by weak bulk perturbations cannot be relied on anymore. In such a case, the Landau energy levels would still lose their degeneracy, but the levels would then become split in a such a way that they cannot be investigated analytically as done here.

\section{The possibility of testing the ISL}\label{sec:V}
As we saw in Sec.\,\ref{sec:III}, the Hall resistivity depends on the gravitational acceleration $g$ for the case of a QHE under the influenced of Earth's gravitational field. Also, as we just saw in the previous section, the broadening of the Landau levels and, hence, the reshaping of the resistivity plateaus depend on the parameter $K$ coming from the gravitational field between the two hemispheres. As such, we might expect that the extent to which the Hall resistivity is affected and the extent to which the Hall plateaus are reshaped would depend, respectively, on the exact form of the gravitational interaction encoded inside $g$ and inside the factor $K$. For the case of the Earth's gravitational field, even though we saw that the correction brought to the quantized Hall resistivity is measurable, it is clear that any different method for measuring $g$ would be more efficient for investigating a possible departure from the ISL of gravity. Let us then examine also the possibility, at least in theory, for such a departure to affect the shape of the dependence of the Hall resistivity with the strength of the magnetic field.

For the case of a Yukawa-like correction to the ISL, we have, according to Eq.\,(\ref{HemisphereNew+Yuk}) of Appendix \ref{A},
\begin{equation}\label{KNew+Yuk}
    K\approx\frac{4\pi G\rho}{3}\left(1-\frac{3 a}{2R}+\frac{\alpha R}{\lambda}e^{-R/\lambda}\right).
\end{equation}
Although the presence of the exponential factor $e^{-R/\lambda}$ in the last term makes the latter exponentially suppressed, the smallness of the ratio $a/R$ suggests that for some special values of the parameters $\alpha$ and $\lambda$, the second and the last terms inside the parentheses might become comparable. Unfortunately, for this to happen, with a separation distance $2a$ between the two hemispheres of the order of a fraction of a millimeter, one needs to have the term $\frac{\alpha}{\lambda} Re^{-\frac{R}{\lambda}}$ of the order of $10^{-4}$. Only for a separation distance between the hemispheres of the order of a fraction of a micrometer, would the term $\frac{\alpha}{\lambda} Re^{-\frac{R}{\lambda}}$ be allowed to be as low as $10^{-6}$. But, then, such a correction would already become too small to make any detectable difference.

For a power-law deviation from the ISL, we have, according to Eq.\,(\ref{Power}), 
\begin{equation}\label{KPower}
K\approx\frac{4\pi G\rho}{3}\left(1-\frac{3 a}{2R}\right)\left(1+\frac{2r_0}{R}\right).
\end{equation}
In contrast to the Yukawa-like correction, the absence of the exponential factor here implies that the allowed range of the distance scale $r_0$ of the deviation to be probed is directly determined by the separation distance between the hemispheres. In fact, a separation distance between the two hemispheres of the order of a micrometer allows one to investigate departures from the ISL down to distance scales of the micrometer too. With such a deviation, and given the order of magnitude of the uncorrected $K$ --- which is $\sim10^{-6}\,$rad$^2$/$s^2$ --- formula (\ref{KPower}) indicates that we need to detect in the QHE the effect of gravity up to the order of $10^{-12}\,$rad$^2$/s$^2$. Now this might seem to be a minute quantity to be detected experimentally, but what the QHE would actually allow to detect is not those individual tiny differences in energies, but the accumulated effect of such available minute energies as the 2DEG gradually occupies those available sub-levels when the magnetic field is gradually decreased. Unfortunately, formula (\ref{LandauGravityFirstOrderInduced}) indicates that such a correction to the term $K$ would affect evenly all the individual orbitals $\ell$ at once. In such a case, the only effect on the 2DEG would be to modify the distance between the available sub-levels without modifying the total available number of these sub-levels. Therefore, the difference between the effect on the QHE of an ISL of gravity and that which deviates from the latter would not be possible to detect in the lab.

\section{summary}\label{sec:VI}
We have examined in this paper the effect of gravity on the QHE. We first examined such an effect when due to the uniform gravitational field of the Earth and derived the correction to the quantized Hall resistivity brought by the field. Plugging in actual experimental values for the various parameters of the setup, we saw that the resulting apparent von Klitzing constant $\tilde{R}_K$ is affected by the gravitational acceleration $g$ in a measurable way. On the other hand, we saw that the gravitational field of the Earth has no effect, neither on the shape of the quantum Hall plateaus nor on the plateau-plateau transitions of the transverse resistivity. We then turned our attention to a different source for the gravitational field. Such a source consists of two very massive hemispheres of the same radius and composition, put very close to each other. We saw that such a configuration does create a nonlinear field of gravity between the two hemispheres that is quadratic in its leading terms. As such, we saw that although the field thus created is weak compared to that of the Earth, the fact that it allows to induce a simple harmonic motion in the 2DEG inside a circular conductor sandwiched between the two hemispheres affects instead the Landau levels of a 2DEG (or any used ion or charged molecule) and their degeneracy. The gap between two adjacent Landau levels shrinks and the energy of each level is split into sub-levels spread over the orbitals. The consequence of this for the QHE is to provide, at each Landau level, sub-levels for the 2DEG to occupy as the magnetic field is decreased. This would then play the same role played by the defects in the sample. That is, as the magnetic field is increased, the gap between adjacent Landau levels increases but the 2DEG is still allowed to occupy the remaining accessible sub-levels within a given principle level $n$. As such, we expect to see a more gentle transition of the Hall resistivity from one plateau to another than the one observed in the absence of the gravitational field, for the sub-levels that arise are all extended (current-carrying) states. Such an effect involves the highly and very actively investigated plateau-plateau transitions of the quantum Hall resistivity. We saw that the technological challenges associated with cryogenics is the only obstacle in actually observing the described effect induced by gravity as it requires extremely low temperatures.

We investigated in detail the issue of the gravitationally induced electric field in the sample. We saw that, in contrast to what an induced electric field does in experiments that aim at testing the equivalence principle, in the case of the QHE such an induced electric field has a beneficial effect. In fact, we saw that all the effects of gravity, whether due to the uniform Earth's gravitational field or the nonlinear field created between the two hemispheres, become amplified. We found, as a result, that the correction to the quantized Hall resistivity increases by a few orders of magnitude. Similarly, we saw that the separation distance between adjacent Landau levels and their splitting become also amplified by the same amount.

Finally, we have examined the possibility for the QHE to play an eventual role in the modern investigations of gravity based on mesoscopic systems \cite{Kiefer} or those based on the use of ultracold neutrons \cite{Kulin,Abele,Biedermann,COWball}. We came to the conclusion that, even though gravity does influence the Hall resistivity, the Landau levels and their degeneracy, and hence also the QHE, the latter cannot really allow to differentiate between an ISL of gravity and another law that deviates from the latter. We have examined in the process both a Yukawa-like deviation from the ISL law and a power-law deviation. Both implied that no distinction could easily be made, either from measuring the minute deviation in the quantized Hall resistivity or from the plot of the variation of the quantum Hall resistivity with the magnetic field strength. Yet, the influence of gravity on the quantum Hall effect has been demonstrated and could thus be used as a means for high precision measurements of the gravitational acceleration $g$ and (once much lower temperatures become easily accessible in the laboratory) of the gravitational constant ${G}$.

We have focused in this paper solely on the integer quantum Hall effect as it is easier to implement in it the effect of the gravitational field. It would be very interesting, though, to investigate also the effect of gravity on the fractional quantum Hall effect (and even other phenomena involving quantum effects such as superconductivity). However, such an investigation requires to take into account also the interaction between the electrons. In addition, given the very small separation between the hemispheres required for our approximations, it is natural to also try to exploit the thin films of graphite \cite{Nature}. We should defer such an investigation to a future work as those tasks remain beyond the scope of the present paper.
\section*{Acknowledgments}
The authors are grateful to the anonymous referee for his/her comments that improved the clarity and quality of our presentation. This work is supported by the Natural Sciences and Engineering Research
Council of Canada (NSERC) Discovery Grant (RGPIN-2017-05388).
\appendix
\section{The gravitational potential inside a full disk of radius $R$ and thickness $a$}\label{A}
\subsection{Newtonian potential $V(r)=-GM/r$.}
Let us denote the Newtonian gravitational potential between the two hemispheres of Sec.\,\ref{sec:III} at any point $x$ away from the center by $V_{H}^N(x)$. Then, we have $V_H^N(x)=V_S^N(x)-V_D^N(x)$. Here, $V_S^N(x)$ is the Newtonian gravitational potential at any distance $x$ from the center of a full sphere of radius $R$ and $V_D^N(x)$ is the Newtonian gravitational potential at the distance $x$ from the center of a full disk of thickness $2a$ and radius $R$. We are going to give here the detailed calculations that yield $V_D^N(x)$. The calculation of $V_S^N(x)$ has been done in detail in Ref.\,\cite{COWball}, so we are only displaying here the final expression of that potential. It is given by,
\begin{equation}\label{SphereNew}
V_S^N(x)=-2\pi G\rho\left(R^2-\frac{x^2}{3}\right).
\end{equation}

The Newtonian gravitational potential at a distance $x$ from the center of a full disk of thickness $2a$, of radius $R$ and of mass density $\rho$, can be found by integrating first the infinitesimal contributions of the mass elements $r\,{\rm d}\phi\,{\rm d}r\,{\rm d}z$ over the thickness $2a$. These contributions should, in turn, be integrated over the whole disk by following the same strategy as the one adopted in Ref.\,\cite{LassBlitzer}. We find,
\begin{align}\label{DiskNew}
    V_D^N(x)&=-4G\rho\int_0^\pi\int_0^{r(\phi)}\int_0^{a}\frac{r}{\sqrt{r^2+z^2}}\,{\rm d}\phi\,{\rm d}r\,{\rm d}z\nonumber\\
    &=-2G\rho\int_0^\pi\Bigg[a\sqrt{r^2(\phi)+a^2}+r^2(\phi)\ln\left(\frac{\sqrt{r^2(\phi)+a^2}+a}{r(\phi)}\right)-a^2\Bigg]{\rm d}\phi.
\end{align}
In the second step we have integrated over the rest of the disk from $r=0$ to $r(\phi)=x\cos\phi+\sqrt{R^2-x^2\sin^2\phi}$ \cite{LassBlitzer}. For small distances $x$ away from the center of the disk and for a small thickness $2a$ of the disk compared to the radius $R$ of the latter, we may, in turn, expand the integrand in powers of the ratios $x/R$ and $a/R$. This allows us then to easily integrate over the variable $\phi$. Keeping only the lower orders in the expansion, we find the following result for the Newtonian gravitational potential inside the full disk at a distance $x$ from its center:
\begin{equation}\label{DiskNewExpansion}
V_D^N(x)=\frac{\pi G\rho a}{R}x^2+\mathcal{O}\left(\frac{x^4}{R^4},\frac{x^2a^2}{R^4}\right)+{\rm const}.
\end{equation}
We have discarded here orders four and higher in $x/R$ and $a/R$. Also, we have gathered all the constant terms inside the last term. Using this result, we can now find the Newtonian gravitational potential between the two hemispheres at a distance $x$ from the center by using Eq.\,(\ref{SphereNew}) and computing $V_H^N(x)=V_S^N(x)-V_D^N(x)$. We thus find the following result:
\begin{equation}\label{HemisphereNew}
V_H^N(x)=\frac{2\pi G\rho}{3}\left(1-\frac{3 a}{2R}\right)x^2+\mathcal{O}\left(\frac{x^4}{R^4},\frac{x^2a^2}{R^4}\right)+{\rm const}.
\end{equation}
\subsection{Yukawa-like correction $V^{\rm Y}(r)=-\frac{1}{r}GM\alpha e^{-r/\lambda}$.}
The gravitational potential inside a full sphere of radius $R$ and mass density $\rho$ due to the Yukawa-like correction term has already been found in Ref.\,\cite{COWball}. The result is
\begin{equation}\label{SphereYuk}
V_S^{Y}(x)
=-2\pi G\rho\Bigg[\alpha\lambda^2\left(2-e^{-\frac{R-x}{\lambda}}-e^{-\frac{R+x}{\lambda}}\right)+\frac{\alpha\lambda^2}{x}\bigg((R+x+\lambda)e^{-\frac{R+x}{\lambda}}-(R-x+\lambda)e^{-\frac{R-x}{\lambda}}\Bigg)\Bigg].
\end{equation}
This potential cannot be of any use in the form given here so we need to expand it in powers of $x/\lambda$. This expansion would be valid provided, of course, that $x<\lambda$. The result is
\begin{align}\label{SphereYuk}
V_S^{Y}(x)
&=e^{-R/\lambda}\left[\frac{2\pi GR\rho\alpha}{3\lambda}x^2+\mathcal{O}\left(\frac{x^4}{R^4}\right)\right]+{\rm const}.
\end{align}
We have kept here also only the terms up to the second order in $x/\lambda$ and gathered all the constant terms inside the last constant term.

Let us now denote by $V_D^{\rm Y}(x)$ the Yukawa-like correction to the gravitational potential at a distance $x$ from the center of a full uniform disk of radius $R$ and of thickness $2a$, having the uniform mass density $\rho$.
Integrating over the disk as done for the Newtonian potential, we get,
\begin{align}\label{DiskYuk1}
    V_D^Y(x)&=-4G\rho\alpha\int_0^\pi\int_0^{r(\phi)}\int_0^{a}\frac{re^{-\sqrt{r^2+z^2}/\lambda}}{\sqrt{r^2+z^2}}\,{\rm d}\phi\,{\rm d}r\,{\rm d}z\nonumber\\[5pt]
    &\approx e^{-\frac{\sqrt{R^2+a^2}}{\lambda}}\left[\frac{\pi G\rho a\alpha}{R}x^2\!+\!\mathcal{O}\!\left(\frac{x^4}{R^4},\frac{x^2a^2}{R^4}\right)\!+\!{\rm const}.\right]
\end{align}
Note that this is the lowest limit of the integral. Indeed, given that the integral in the first line of Eq.\,(\ref{DiskYuk1}) does not admit any analytical expression, we have to approximate the exponential inside the integrand in order to get a rough estimate of such an integral. Then we integrated over the whole disk as we did above for the Newtonian potential by assuming that $x\ll R$. Owing to the smallness of $a$, we clearly see that the contribution of this lower limit of $V_D^{Y}(x)$ is very small. Finally, then, using Eqs.\,(\ref{HemisphereNew}) and (\ref{SphereYuk}), the gravitational potential $V_H^{N+Y}(x)$ that takes into account the Yukawa-like correction at any distance $x$ from the center between two hemispheres is found, up to the first two lowest orders, as follows:
\begin{equation}\label{HemisphereNew+Yuk}
    V_H^{N+Y}(x)\approx V_H^N(x)+V_S^{Y}(x)\approx\frac{2\pi G\rho}{3}\left(1-\frac{3 a}{2R}+\frac{\alpha R}{\lambda}e^{-R/\lambda}\right)x^2+{\rm const}.
\end{equation}
Note that this approximation is valid provided that $x^4/R^4\ll \alpha R/\lambda e^{R/\lambda}$. For this to be the case, one needs of course to choose the allowed range for $x$ according to one's estimate of $\lambda$ and $\alpha$.
\subsection{Power-law correction $V^{n}(r)=-GM\,r_0^n/r^{n+1}$.}
Let us denote by $V_D^{n}(x)$ the power-law correction to the gravitational potential at a distance $x$ from the center of a full uniform disk of radius $R$ and thickness $2a$, having the uniform mass density $\rho$.
\subsubsection{Case $n=1$.}
Only for the case $n=1$ do we have a converging potential inside the sphere \cite{COWball}. Therefore, we only focus here on that case. The gravitational potential correction at a distance $x$ from the center of a sphere of radius $R$ and mass density $\rho$ for such a case has already been computed in detail in Ref.\,\cite{COWball}. Therefore, we just pick up here the result found there and expand in powers of $x/R$. We find,
\begin{align}\label{SpherePL1}
    V_S^{n=1}(x)&=-2\pi G\rho r_0\left[R+\frac{R^2-x^2}{2x}\ln\frac{R+x}{R-x}\right]\nonumber\\
    &=-4\pi G\rho r_0R\left(1-\frac{x^2}{3R^2}\right)+\mathcal{O}\left(\frac{x^4}{R^4}\right)+{\rm const}.
\end{align}
On the other hand, we should integrate over the disk of radius $R$ and thickness $2a$ as done for the Newtonian potential. We get,
\begin{align}\label{DiskPL1}
V_D^{n=1}(x)&=-4G\rho r_0\int_0^\pi\int_0^{r(\phi)}\int_0^{a}\,\frac{r}{r^2+z^2}\,{\rm d}\phi\,{\rm d}r\,{\rm d}z\nonumber\\[5pt]
&=-2G\rho r_0 \int_0^\pi\Bigg[a\ln\left(\frac{a^2+r^2(\phi)}{a^2}\right)+r(\phi)\tan^{-1}\left(\frac{a}{r(\phi)}\right)\Bigg]{\rm d}\phi\nonumber\\[5pt]
&=\frac{2\pi G\rho r_0a}{R^2}\,x^2+\mathcal{O}\left(\frac{x^4}{R^4},\frac{x^2a^2}{R^4}\right)+{\rm const}.
\end{align}
As the integral in the second line does not again admit any analytical solution, the best we could do is expand the integrand in powers of $x/R$ and $a/R$ and then integrate for the single variable $\phi$. The above result is thus  valid only for $x\ll R$ as well as $a\ll R$.
Therefore, the total gravitational potential between the two hemispheres reads,
\begin{equation}\label{Power}
V_H^{N+(n=1)}(x)=V_H^{N}+V_S^{n=1}-V_D^{n=1}\approx\frac{2\pi G\rho}{3}\left(1-\frac{3 a}{2R}\right)\left(1+\frac{2r_0}{R}\right)x^2+{\rm const}.
\end{equation}

\end{document}